\documentclass{article}

\usepackage{arxiv}
\usepackage[utf8]{inputenc} % allow utf-8 input
\usepackage[T1]{fontenc}    % use 8-bit T1 fonts
\usepackage{hyperref}       % hyperlinks
\usepackage{url}            % simple URL typesetting
\usepackage{booktabs}       % professional-quality tables
\usepackage{amsfonts}       % blackboard math symbols
\usepackage{nicefrac}       % compact symbols for 1/2, etc.
\usepackage{microtype}      % microtypography
\usepackage{lipsum}
\usepackage{amssymb}
\usepackage{amsmath}
\usepackage{bm}
\usepackage{stmaryrd}
\usepackage{placeins}
\usepackage[lofdepth,lotdepth]{subfig}
\usepackage{graphicx}
\graphicspath{{./figs/}}
\usepackage[titletoc]{appendix}
\newcommand{\db}[1]{\ensuremath{\llbracket #1 \, \rrbracket}}

\title{The Piano Inpainting Application}

\author{
  Ga\"etan Hadjeres\thanks{Equal contribution} \\
  Sony Computer Science Laboratories\\
  Paris, France\\
  \texttt{gaetan.hadjeres@sony.com} \\
   \And
  Léopold Crestel${}^*$\\
  Sony Computer Science Laboratories\\
  Paris, France\\
  \texttt{leopold.crestel@sony.com} \\
}
\date{}
\begin{document}
\maketitle

\begin{abstract}
Autoregressive models are now capable of generating high-quality minute-long expressive \textit{MIDI} piano performances. 
Even though this progress suggests new tools to assist music composition, we observe that generative algorithms are still not widely used by artists due to the limited control they offer, prohibitive inference times or the lack of integration within musicians' workflows.
In this work, we present the Piano Inpainting Application (PIA), a generative model focused on ``inpainting'' piano performances,  as we believe that this elementary operation (restoring missing parts of a piano performance) encourages human-machine interaction and opens up new ways to approach music composition. 
Our approach relies on an encoder-decoder Linear Transformer architecture trained on a novel representation for \textit{MIDI} piano performances termed \emph{Structured MIDI Encoding}. 
By uncovering an interesting synergy between Linear Transformers and our inpainting task, we are able to efficiently inpaint contiguous regions of a piano performance, which makes our model suitable for interactive and responsive A.I.-assisted composition.
Finally, we introduce our freely-available \emph{Ableton Live} PIA plugin, which allows musicians to smoothly generate or modify any \textit{MIDI} clip using PIA within a widely-used professional Digital Audio Workstation. 
\end{abstract}

% keywords can be removed

\section{Introduction}
\label{sec:introduction}
% motivation
With the recent advances in autoregressive models \cite{vaswani2017attention,katharopoulos2020transformers} and increasing computational resources, convincing minute-long piano performances with fine details about velocity and micro-timing can be automatically generated \cite{huang2018music,musenet,huang2020pop}, opening the way to cutting edge creative assisting tools for composition.
However, unconstrained generation can be of very limited use for composers, and engaging interactions between a user and the algorithm has been identified as a major requirement towards A.I.-assisted music composition.
Although it proved to be challenging to develop meaningful controls for deep learning models, many recent works proposed various modes of interaction for music generation beyond the usual continuation of a priming section \cite{musenet,huang2018music,huang2020pop,hsiao2021compound}: harmonization \cite{hadjeres2017deepbach,huang2018music}, variation of an input sequences \cite{hadjeres2017glsr,Hadjeres2020}, interpolation between two fragments \cite{magentastudio}, or mapping of a musical gesture to a performance \cite{Donahue2019}.

\begin{figure*}[t]
\centering
\includegraphics[scale=0.25]{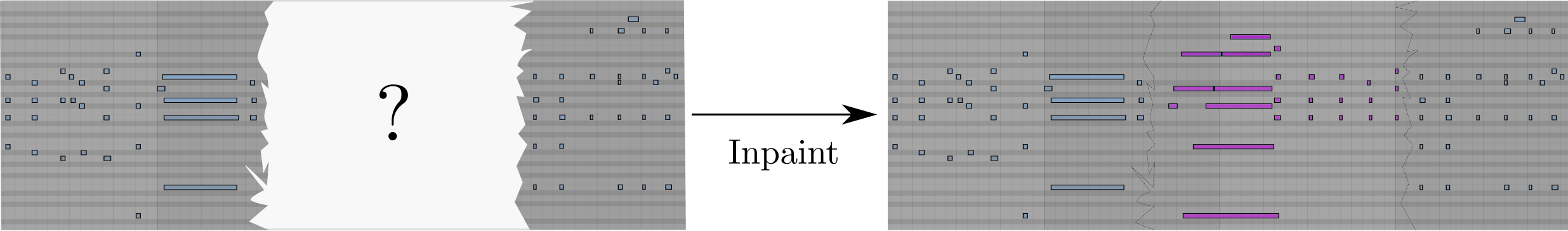}
\caption{
The piano inpainting task consists in restoring an incomplete \textit{MIDI} performance.
}
\label{fig:inpainting}
\end{figure*}

In this work, we choose to focus on the piano performance inpainting task, which we can define as learning how to restore a partially blanked-out piano performance as depicted in Fig.~\ref{fig:inpainting}.
The reason for considering the piano inpainting task as our task of interest is that this operation encompasses many previously cited generation schemes (unconditional generation, continuation of a priming sequence) but also favors an iterative compositional process.

This task has been extensively studied for image generation \cite{DBLP:journals/corr/abs-1909-06399}.
However, we note that in the symbolic music domain only few approaches were proposed and that these methods only operated on quantized scores:
Bach chorales inpainting using Gibbs sampling where the inpainted elements are initialised randomly and resampled until convergence was proposed in \cite{hadjeres2017deepbach},
Bach chorales melodies inpainting using RNNs was developed in \cite{hadjeres2020anticipation}, 
while \cite{Pati2019} proposed a VAE-based solution for variable-length folk music melody inpainting.

Extending these results to the more complex task of generating expressive piano performances is not straightforward, as the previous results largely exploit the fixed structure of the data representation to provide exact information about the location of the constraints (for example, in \cite{hadjeres2020anticipation}, the fifth beat of the soprano voice of a Bach chorale will always have the same position in the encoding sequence).
However, recent works in piano performance generation rely on data representations inspired by the event-based encoding of the \textit{MIDI} format \cite{huang2018music,musenet,donahue2019lakhnes,huang2020pop}, whose structure depends on the content of the encoded performance. 
However, if event-based encoding proved to be sufficient when working with autoregressive causal models, their relative lack of structure induces severe challenges for the inpainting task.

These recent models for expressive piano generation relied on the latest advances involving the Transformer architecture \cite{vaswani2017attention} and their variants \cite{transfoxl,Shaw2018,child2019generating,katharopoulos2020transformers} to improve generation quality but seldom considered the inference speed as a key element. However, if we require these A.I. models to be used as a creation tool involving a human operator, inference speed becomes a major concern as it makes the interaction fluid and enhances the user experience. 

In this paper, we propose the Piano Inpainting Application (PIA), a complete system for inpainting a piano
performance. This system is composed of a backend based on an optimized Linear Transformer \cite{katharopoulos2020transformers} encoder-decoder architecture together with a client which comes as a \emph{Ableton Live} plugin written in \emph{Max for Live}. 
This allows our inpainting model to be seamlessly integrated in a popular Digital Audio Workstation (DAW), enriches this DAW rich plugin ecosystem, while making it accessible to a wide audience. 
We introduce a novel representation for expressive piano performances termed \emph{Structured MIDI Encoding} which is better suited for the piano inpainting task than previously-used representations.
By taking advantage of the two modes of inference of the Linear
Transformer, we propose a fast inference scheme to inpaint contiguous
regions that is key to the reactivity of PIA.

% Paper structure
The remainder of this paper is organized as follows:
Section~\ref{sec:model} introduces the \emph{Structured MIDI
  Encoding} and describes our proposed model
architecture. Experiments are detailed in Sect.~\ref{sec:experiments}, in which we present the dataset we used,
implementation details and showcase unsuspected applications, beyond
the regeneration of contiguous regions. The PIA \emph{Ableton
  Live} plugin is briefly discussed in Sect.~\ref{sec:pia-plugin}, followed by 
a more general discussion on related works in Sect.~\ref{sec:related-works}.%, and a conclusion is provided in Section \ref{sec:conclusion}.

\section{Model}
\label{sec:model}
We first present  in Sect.~\ref{sec:backgr-inpa-with} the framing of the inpainting task using autoregressive models as exposed in \cite{hadjeres2020anticipation} before introducing in Sect.~\ref{sec:ss:encoding} a novel representation for expressive piano performances that we called \emph{structured MIDI encoding}. This representation is general, well-suited when training generative models, and its regular structure allows us to design an efficient encoder-decoder architecture to perform piano inpainting, which we present in Sect.~\ref{sec:ss:transformers}.
We finally describe in Sect.~\ref{sec:efficient-inference} a fast inference scheme to perform inpainting of contiguous regions.

\subsection{Background: inpainting with autoregressive models}
\label{sec:backgr-inpa-with}
The Anticipation-RNN framework \cite{hadjeres2020anticipation} casts
the inpainting task as a conditional generation problem. Suppose we are given a dataset of
discrete sequences $\mathbf{x} = (x_t)_{t \in \db{T}}$ where
$x_t \in \mathcal{A}$ belongs to a discrete alphabet $\mathcal{A}$ and $T$ denotes the length of
the sequences, which are assumed to be samples from a distribution $p(\mathbf{x})$\footnote{We use the shorthand $\db{d}$ to denote $\db{1, \dots, d}$}. Given a set of positional constraints $\mathcal{C} :=  \{(t, c_t) \in T \times
\mathcal{A})\}$, 
 the objective is to generate sequences $\mathbf{x}$ enforcing these
 constraints, i.e. such that $x_t =
 c_t$ for all $(t, c_t)$ in $\mathcal{C}$ with the correct
 probabilities. More precisely, this amounts to sampling from the subset
 of sequences enforcing the constraints $\mathcal{C}$ with
 $p(\mathbf{x}|\mathcal{C})$ while keeping the relative probabilities
 (as defined by $p(\mathbf{x})$) intact. Note that by definition,
 $p(\mathbf{x}|\mathcal{C})=0$ for sequences that violate one of the
 constraints in $\mathcal{C}$, so we have
 $p(\mathbf{x}|\mathcal{C}) \propto p(\mathbf{x})$ on the subset of
 valid sequences.
 
% Problem
However, enforcing constraints
when generating time series with an autoregressive model
$p(x_t|\mathbf{x}_{<t})$ is challenging and rejection sampling can become
highly inefficient with a high number of positional constraints.
% Solution
The solution proposed in \cite{hadjeres2020anticipation} consists in
summarizing the set of constraints $\mathcal{C}$ with
an autoregressive model going backwards and then to properly condition
the generation of the sequence $\mathbf{x}$ using such information
about the future.

\begin{figure}
\centering
\includegraphics[scale=0.6]{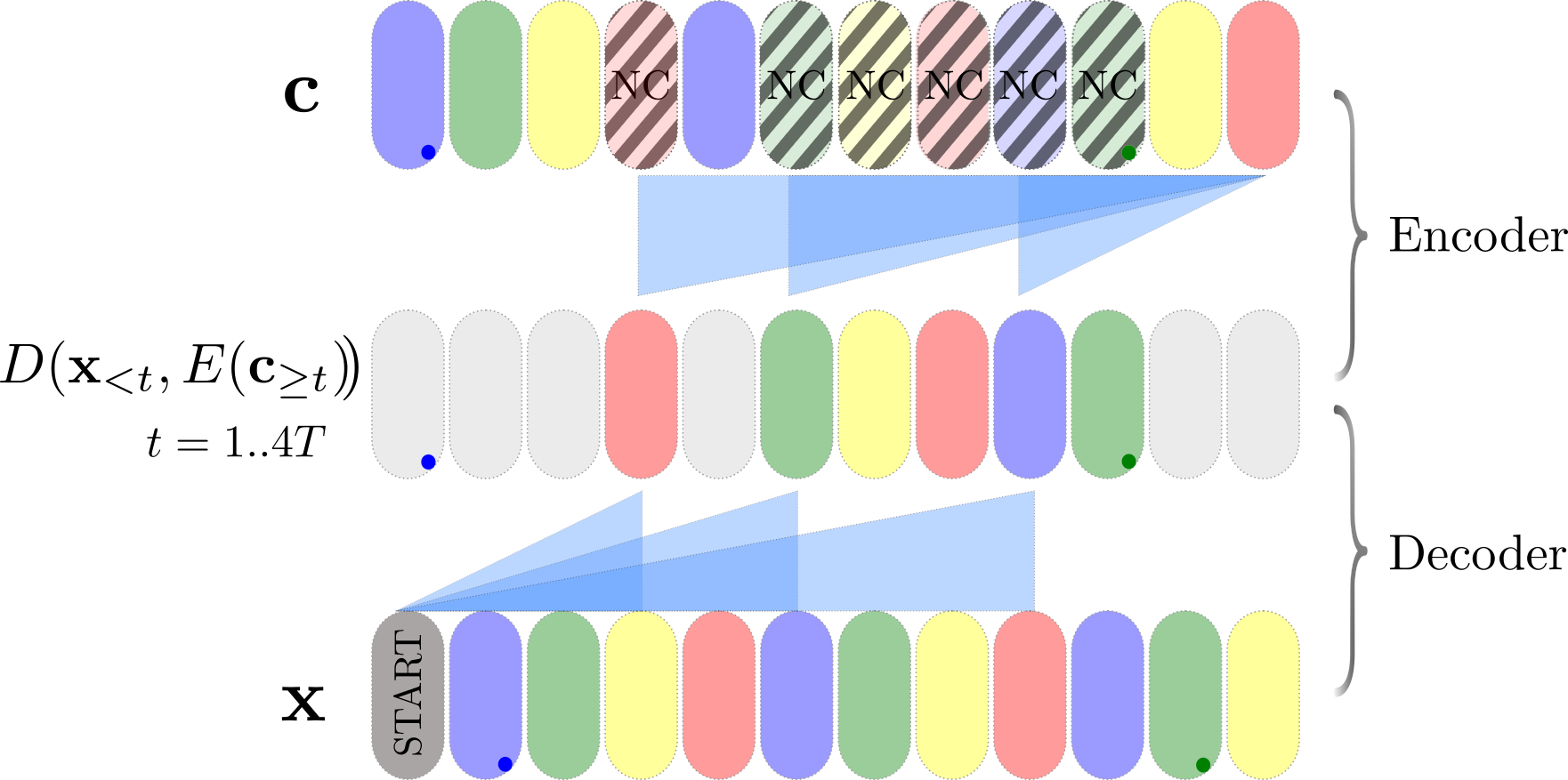}
\caption{
  Schematization of our proposed masking scheme. Output tokens (middle
  row) can only attend to $\mathbf{x}$ and $\mathbf{c}$ tokens
  delimited by blue triangles;  $\mathbf{c}$ is a masked version of
  $\mathbf{x}$. For ease of implementation, a
  $\mathbf{START}$ token is appended before $\mathbf{x}$. Our loss
  function only  involves
  predictions over Non Constrained tokens.
} 
\label{fig:structured-transformer}
\end{figure}

% Details
In practice, a set of constraints $\mathcal{C}$ is represented as a sequence of
length $T$ by adding a $\mathrm{NC}$ (No Constraint) token:
$\mathbf{c} = (c_t)_{t \in \db{T}}$, where $c_t = \mathrm{NC}$ when
there is no constraint at location $t$ in $\mathcal{C}$.
The sequence $\mathbf{c}$ is fed to an anti-causal autoregressive
encoder $E$, whose output sequence $(E(\mathbf{c}_{\geq t}))_{t \in \db{T}}$ represents at each time step a summary of the constraints occurring at future times. 
It is used to condition a causal decoder $D$ which approximates
$p(\mathbf{x}|\mathcal{C})$ in an autoregressive way  (as depicted in 
Fig.~\ref{fig:structured-transformer}):
\begin{equation}
  \label{eq:arnn}
  D(\mathbf{x}_{<t}, E(\mathbf{c}_{\geq t})) \approx p(x_t|\mathbf{x}_{<t},\mathcal{C}).
\end{equation}
Training is done by sampling
 constraint locations and minimizing a cross-entropy loss. Inference
 is done by first computing $(E(\mathbf{c}_{\geq t}))_{t \in \db{T}}$
 and then successively sampling $x_t \sim  D(\mathbf{x}_{<t}, E(\mathbf{c}_{\geq t}))$.

When working with quantized scores, this approach proved to be effective at modelling the correct
probability distribution while enforcing the constraints and provided
an efficient sampling scheme requiring only $O(2T)$ RNN calls.
% This approach allows to place constraints at any position in the sequence, which leads to a very flexible inpainting process where any sub-part of an initial sequence can be re-generated at will.
% However, the number and position of the inpainted tokens is fixed which is an important limitation.

\subsection{Structured \textit{MIDI} encoding}
\label{sec:ss:encoding}
Directly applying the method described in Sect.~\ref{sec:backgr-inpa-with} to piano performances proved to be challenging.
Indeed, one may notice that defining constraint sequences was only possible in \cite{hadjeres2020anticipation} due to the particular structure of the data that the authors considered: when working with quantized scores, there is a direct link between the location $t$ of the constraint within the sequence and the actual onset of the note in the final score.
However, piano performance encoding found in the literature \cite{huang2018music,musenet,donahue2019lakhnes,huang2020pop} do not have an unequivocal relation between an information in the performance and its position in the sequence of tokens representing the performance, which makes the definition of musically-relevant constraints impossible.
(for example, the same notes played either as a chord or as an arpeggio will be represented with a different number of tokens, and the pitch information of each note will be placed at different positions between the two cases, even though there is the same number of notes).

\begin{figure}[t]
\centering
\includegraphics[scale=0.48]{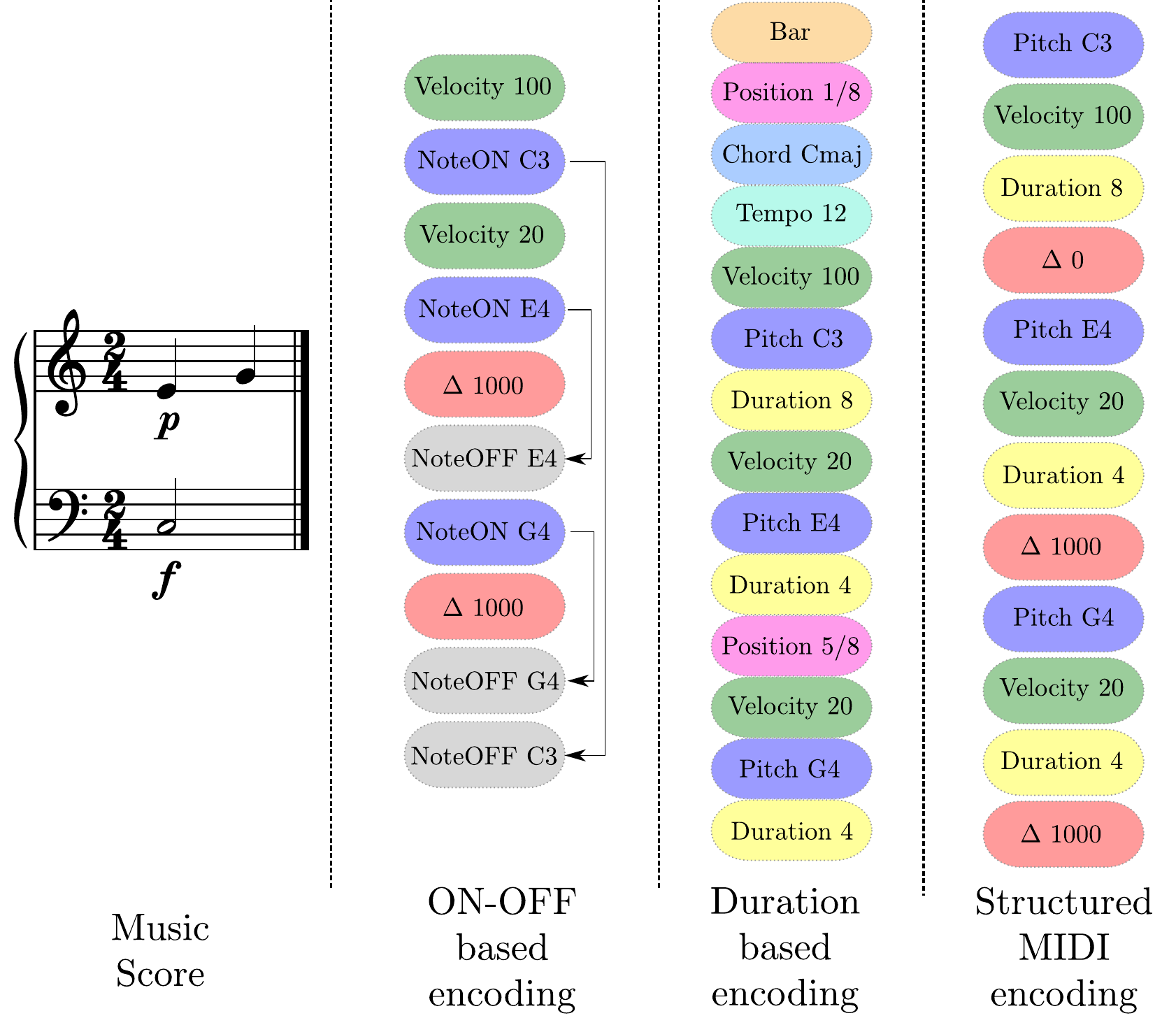}
\caption{
Comparison of three piano performance encodings corresponding to the
music sheet showed on the left.
The On-Off representation used in \cite{huang2018music}  exhibits a
complex nested structure while the 
Duration-based representation from \cite{huang2020pop}  is still less 
structured compared to our proposed \emph{Structured MIDI encoding}.
}
\label{fig:structured-encoding}
\end{figure}
% former caption:
% The On-Off representation used in \cite{huang2018music} is close to the original \textit{MIDI} encoding, but exhibit a relatively complex nested structure that a model has to understand in order to generate meaningful sequences.
% Duration-based representation used in \cite{huang2020pop} simplify time representation, but still lack structure compared with our representation.
% definition
In this work, we consider piano performances encoded as a sequence of notes, each
note being characterized by four attributes or channels:
\textit{Pitch}, \textit{Velocity}, \textit{Duration} and \textit{Time
  Shift}.
The \textit{Time Shift} channel encodes the time elapsed between the onset of the current note and the onset of the next one.
Formally, we encode piano performances as a structured sequence $\mathbf{x} =
(x_t)_{t \in \db{4T}}$ , where $T$ is the number of notes
of the piano performance. For each slice $(x_{4i},x_{4i+1}, x_{4i+2},
x_{4i+3}):=(x^p_i, x^v_i, x^d_i, x^{ts}_i)$ with $i \in \db{T}$, its
entries correspond respectively to the pitch of the $i^{th}$ note, its
velocity, its duration and to the time shift between the $i^{\text{th}}$ and $(i+1)^{\text{th}}$ notes. 
Each slice can be seen as a tuple in $\mathcal{A}_p \times  \mathcal{A}_v \times \mathcal{A}_d \times \mathcal{A}_{ts}$, meaning that each of its entries belong to a domain-specific finite alphabet $\mathcal{A}_*$. 
Our representation presents similarities with the duration-based representation proposed in \cite{huang2020pop}, but it is more general (works with any \textit{MIDI} file without annotation about bars or tempo) and is more structured.
Notably, we introduce a token corresponding to 0 milliseconds in the time shift alphabet, which is used when several notes occur at the same time. 
This is crucial to guarantee the structure of the representation as visualized in Fig.~\ref{fig:structured-encoding}. Our modeling choices for the alphabets $\mathcal{A}_*$ are discussed in Sect.~\ref{sec:adapt-quant-durat}.

% adaptive quantisation scheme later on 

% structured because duration based and time-shift 0
An important property of our proposed representation is that the data always keeps the same structure independently from the content.
As a consequence, for each token index $t \leq 4T$, we know exactly the nature of the information (pitch, velocity,
duration or time shift) it represents and where the  $i^{\text{th}}$ played note is located.
This differs from other piano performance representations, whose structure vary depending on the content.
We discuss the differences between our proposed structured
representation and other \textit{MIDI}-like representations in
Sect.~\ref{sec:ss:midi-encoding-rel}.
\subsection{Efficient Linear Transformer for piano inpainting}
\label{sec:ss:transformers}
Using the \emph{Structure MIDI encoding} for piano performances, we
propose to rely on an encoder-decoder Transformer architecture
\cite{vaswani2017attention} to model Eq.~(\ref{eq:arnn}). We take
advantage of the particular structure of our data representation to
propose an efficient encoder-decoder architecture based on the
recently-proposed Linear Transformer
\cite{katharopoulos2020transformers} in Sect.~\ref{sec:effic-mask-scheme}. The regularity of our data
representation allows us to design factorized positional embeddings,
which we present in Sect.~\ref{sec:fact-posit-embedd}. We conclude in
Sect.~\ref{sec:efficient-inference} by
proposing an efficient inference  that strictly outperforms the
approach from \cite{hadjeres2020anticipation} when the
inpainted region is contiguous.

%which we adapt to the inpainting task we are trying to address.
%More precisely, our implementation allows to enforce pre-defined constraints in the generation of a time series, while conditioning the sampling at any position under these constraints.
%Notably, while relying on purely autoregressive models, we manage to condition the sampling of any token under both the past tokens and tokens representing constraints occurring at a later position in the sequence.
%This is mainly achieved by implementing an anti-causal Transformer on a sequence representing the constraint, as we describe in the next paragraph.
%The result is a flexible model which allows to efficiently enforce constraints at any position in the sequence. 
%We also propose a fast inference scheme for inpainting tasks based on Linear Transformer \cite{katharopoulos2020transformers}.

\subsubsection{Efficient masking scheme}
\label{sec:effic-mask-scheme}
% Implementation with Transformers
The encoder $E$ and the decoder $D$ can be implemented by any autoregressive model.

In this work, we use Linear Transformers
\cite{katharopoulos2020transformers} as they provide a way to overcome the
$O(T^2)$ complexity of computing the self-attention, which allows us
to consider longer sequences, and we rely on a standard
encoder-decoder architecture with transformers as proposed in
\cite{vaswani2017attention}.
In order to enforce the particular dependency structure displayed in
Eq.~(\ref{eq:arnn}), we propose to use anti-causal masks on the
encoder part, and causal masks on the decoder part. Thanks to the
alignment between the sequence of constraints $\mathbf{c}$ and the
input sequence $x$, cross-attention can be simplified with a much
simpler operation since at location $t$, the decoder only needs
attend to $E(\mathbf{c}_{\geq t})$. Such operation is equivalent
to masking out all off-diagonal terms of the cross-attention matrix in
the standard setting.
Compared with \cite{hadjeres2020anticipation}, replacing RNNs with
Linear Transformers allows to compute predictions for all timesteps in parallel.

\subsubsection{Factorized positional embeddings}
\label{sec:fact-posit-embedd}
Positional embeddings are a key component in Transformer architectures
as they provide a way to convey information about the
sequential ordering of the input. Usually, information about the
index $t$ in the sequence is provided through the use of sinusoidal
positional embeddings \cite{vaswani2017attention}. In our case, due to
the regular structure of the Structured MIDI encoding, each index $t$
indicates two different kinds of information: the channel index of
the corresponding token $t \, \mathrm{mod} \, 4$, and the index of the note
it belongs to $\left \lfloor{t/4}\right \rfloor$.

We also decide to introduce \emph{elapsed time embeddings} to 
indicate at each location $t$ the true elapsed time in seconds since
the beginning of the sequence (this is obtained by
aggregating all time shifts). We note that the elapsed-time embedding allows the decoder to know the duration of the gap that has to be inpainted.
Such embedding seems to have been used in \cite{musenet}, but probably in a slightly different way: in our case, both the positional embedding $\left \lfloor{t/4}\right \rfloor$ and the elapsed time embedding are obtained via the sinusoidal embeddings from \cite{vaswani2017attention}.
Details are provided in Appendix.~\ref{sec:app:embeddings}.

\subsubsection{Efficient inference scheme}
\label{sec:efficient-inference}
% Mixed parallel and recursive inference in Linear Transformer
Reactivity is a major concern in our application.
Therefore, we rely on the Linear Transformer
\cite{katharopoulos2020transformers} which can offer a faster inference
time compared with vanilla Transformers. Indeed, inference in Linear
Transformers can be also be performed in a recurrent manner (as in a
Recurrent Neural Network), by computing a single time step at a time
and updating a hidden state. This helps saving resources
when performing autoregressive generation since only necessary
information is computed at each time step.

In our implementation, we take advantage of the two modes of inference
offered by the Linear Transformer to propose an efficient inference
scheme when the inpainted region is contiguous: we first compute in
parallel the output of the encoder and the hidden state needed to
compute the first token of the inpainted region. After this, we can
use the RNN-like inference mode to sample each token of the inpainted
region in an autoregressive way.

This observation allows our method to perform strictly better than the RNN-based
method from \cite{hadjeres2020anticipation}, as the complexity of the
inpainting task for contiguous region is $O(L)$ instead of $O(T)$,
where $L \leq T$ is the length of the region to inpaint (note that in
practice we often have $L \ll T$). This implies that the generation time
is irrespective of the location of the inpainted region within the
sequence. Speedups obtained by using the RNN inference mode instead of
the traditional parallel computation are discussed in \cite{katharopoulos2020transformers}.

\section{Experiments}
\label{sec:experiments}
In this section, we present  the dataset we used in
Sect.~\ref{sec:dataset} and precise implementation and training details in
Sect.~\ref{sec:impl-deta}. We then showcase in
Sect.~\ref{sec:music-relev-appl} many generation schemes
granted by our proposed architecture. 
All the results and generations can be found on the companion website of this article \footnote{\href{https://ghadjeres.github.io/piano-inpainting-application/}{https://ghadjeres.github.io/piano-inpainting-application/}}.

\subsection{Dataset}
\label{sec:dataset}
We use the \textit{GiantMIDI-Piano} dataset
\cite{kong2020giantmidi}%\footnote{\href{https://github.com/bytedance/GiantMIDI-Piano}{https://github.com/bytedance/GiantMIDI-Piano}}
, which includes 10,854 piano performances written by 2,786 composer
transcribed  using \cite{kong2020high} from live recordings and encoded in the MIDI format. 
We use a ($90 / 10  / 0$) split for the dataset, by assigning one file every ten to the validation set.

\subsubsection{Data processing}
\label{sec:data-processing}
We process chunks of $1024$ notes, which corresponds to $4096$ tokens
using our Structured MIDI representation introduced in
Sect.~\ref{sec:ss:encoding}. In average, a chunk of $1024$ notes represents at least a minute-long performance.
We use the following data augmentations:
\begin{itemize}
\item time dilation, as a multiplicative factor, uniformly sampled between $0.9$ and $1.1$, applied to values in seconds,
\item velocity shifting, as an additive term on velocity tokens, sampled from a uniform distribution defined on $\db{-20,\dots,20}$,  
\item transposition, as an additive term added to pitches, sampled from a uniform distribution on $\db{-6,\dots,6}$.
\end{itemize}

\subsubsection{Adaptive quantization for continuous values}
\label{sec:adapt-quant-durat}
 We adopt an adaptive quantization scheme for duration and time shift
 events similar to what is proposed in \cite{donahue2019lakhnes}.
 The objective is to discretize continuous duration and time shift
 values efficiently to minimize both loss of information and the
 overall size of their respective alphabets. This results in an almost
 imperceptible discretization. Exact values are provided in appendix \ref{sec:app:encoding_alphabebet}.

\subsection{Implementation details}
\label{sec:impl-deta}

\subsubsection{Improvements to the Linear Transformer architecture and
  model parameters}
\label{sec:impr-line-transf-1}
% tricks
We made two substantial changes to the Linear Transformer from
\cite{katharopoulos2020transformers} that proved to make model
training more stable and improve expressivity.
We first modified how the residual branches and non-residual branches
of all layers
are merged. Following  \cite{parisotto2020stabilizing}, we replaced
the usual addition of the two branches by a GRU-type gating mechanism.
Secondly, following \cite{parisotto2020stabilizing,xiong2020layer}, we
put layer normalization in the residual branches and add an additional
normalization layer before the final softmax layers.

% parameters
Both encoder and decoder are Linear Transformers with $(1 +
\mathrm{elu(x)})$ feature maps, 8 heads of dimension 64 each and
$\mathrm{gelu(x)}$ activations and feed-forward layers of dimension
1024. The encoder has 4 layers and the decoder has 8 layers. Dropout
with value 0.1 is applied.

% output
The model outputs are 
projected using channel-dependant projection heads before computing a
softmax, so that each prediction is done in the correct alphabet (pitch, velocity,
duration or time shift), see Sect.~\ref{sec:ss:encoding} and \ref{sec:adapt-quant-durat}.
% loss
The model is trained by minimizing a cross-entropy loss, computed only for tokens of the output sequence corresponding to
Non-Constrained tokens in the constraint sequence. This is opposed to
the approach of \cite{hadjeres2020anticipation} where the tokens with
positional constraints were predicted as well.
% sampling
At generation, we use top-p sampling with $p = 0.95$  \cite{holtzman2019curious}.

\subsubsection{Constraints at training time}
Depending on the targeted application, different strategies can be
implemented to create the constraint sequences at training time as
discussed in Sect.~\ref{sec:music-relev-appl}.
In our implementation, we remove a whole slice (from $4i$ to $4i + 3$,
$i \leq T$) with probability $p$, where $p$ is a ratio uniformly chosen between $0.5$ and $1$.
This strategy is adapted to the task of filling missing notes in a sequence, but would be, for instance, less adapted to re-sampling a particular information. 
In that later case, masking only the missing information during training would probably be a better strategy.

\subsection{Sampling strategies}
\label{sec:music-relev-appl}
A strength of our model is that it allows many musically-relevant
applications to be performed without the need to rely on different
specific models.

\subsubsection{Unconditional generation}
Our proposed architecture can be used to generate piano performances
from scratch, with no priming or ending section defining a musical
context. In this case, the model constraint
sequence $\mathbf{c}$ is only filled with $\mathbf{NC}$ tokens. We can
use the efficient inference scheme of
Sect.~\ref{sec:efficient-inference} so that the overall complexity of
unconditional generation with our inpainting model is on par with a
causal Linear Transformer trained specifically for unconditional
generation.
Generation quality in this restricted setting is close from other
dedicated approaches \cite{musenet,huang2018music}, despite the fact
that we considered a much smaller model trained on a significantly smaller dataset.
In the generated performances, we observe a real sense of tempo, a
capacity to coherently develop patterns, a wide variety of piano
textures and recognizable musical styles. If we have a feeling of
musical direction, the high-level structure of our generated pieces
may not be as clear as what can be heard in some examples from
\cite{musenet}.
However, we do not perceive this as a major shortcoming since our
proposed approach to composition is through iterative refinements
involving a human operator. In such a setting, it is up to the user to
make decisions about the high-level structure, which can be easily
achieved using our proposed \emph{Ableton Live} PIA plugin presented
in Sect.~\ref{sec:pia-plugin}.

\subsubsection{Inpainting of contiguous regions}
\label{sec:inpa-cont-regi}
Inpainting of a contiguous region (see Fig.~\ref{fig:inpainting}) can be done with the sampling scheme
proposed in Sect.~\ref{sec:efficient-inference}; such setting includes the possibility to continue a priming sequence.
By relying on the elapsed-time embedding and by appropriately choosing
the number of notes within the region-to-be-inpainted, we can specify
both the duration of the region and its note density. This
offers interesting possibilities for a user to control the generated
material. Since selecting a contiguous region is easily done in a
Digital Audio Workstation (DAW), this suggests the DAW integration of
PIA presented in Sect.~\ref{sec:pia-plugin}.

We note that PIA is able to seamlessly stitch together two regions,
and that the desired length for the regenerated region is well
respected. The proposals made by PIA are varied and  take
into account ideas from the context (both from the preceding and
succeeding regions);
% In this section we present the performances of our model for inpainting piano performances.
% We consider a priming section of \textbf{256 NOTES ???} and an ending section of \textbf{256 NOTES ???} and the model has to fill a gap of \textbf{512 NOTES ???}.
% The results produced by our model can be found on the companion website of this paper \footnote{\href{SITE}{SITE}}.

% As a comparison

% Ablation ?? 
% Remove encoders
% Unstructured rep

% Pros: can inpaint notes where you want in the sequence (not conjoint, priming and ending of various sizes)
% pros/cons: duration of inpainted segment is fixed
% cons: number of notes in an inpainted segment is predefined.

\subsubsection{Musically-relevant applications of piano inpainting}
\label{sec:resampl-music-attr}
Our framing of the inpainting task combined with the use of the
Structured MIDI encoding allows us to perform more general operations
than the inpainting of contiguous region described in
Sect.~\ref{sec:inpa-cont-regi} and displayed in
Fig.~\ref{fig:inpainting}.
Indeed, since our representation treats pitch, velocity and note
duration as different tokens, it is possible to perform inpainting
only on some attributes, while leaving the others unaltered. We
identify two interesting applications: ``velocification'' and ``pitchification''.

The former application consists in only regenerating velocities and durations so
that it
becomes possible to ``humanize'' a score or create playing style
variations. This can be particularly relevant when such information is
missing or tedious to input manually. Concrete use cases for this task
are: when one is using  a MIDI keyboard, it might be hard to precisely
record velocities; when inputing a melody in a DAW, it takes time to
also precisely set the duration and the velocity of each note; when
importing a score as MIDI, there may be no information about
velocities so that all notes are assigned to the same velocity.

The latter application consists in doing the opposite: only pitches
are generated, 
 while keeping the original rhythm and the same expressivity
 parameters. It can be a way to further condition PIA generations to
 strictly follow a predefined rhythm,
 but could also lead to creative usages: it is possible for instance
 to record a musical gesture on the piano without any concern about
 the notes being played and to let PIA change this proposal into an
 actual piano piece. Such a creative use bears similarities with the
 approach from \cite{Donahue2019} and the modelling task considered in
 \cite{DBLP:journals/corr/Walder16}.
 
If such applications could be easily performed using a dedicated seq2seq
model where only the relevant attributes in the Structured MIDI
encoding are kept, we find interesting to be able to perform all these
operations with a single model. Furthermore, our model allows to
perform these attribute inpainting tasks only on user-defined regions
of the piano performance.

\begin{figure}[h]
    \centering
    \includegraphics[scale=0.4]{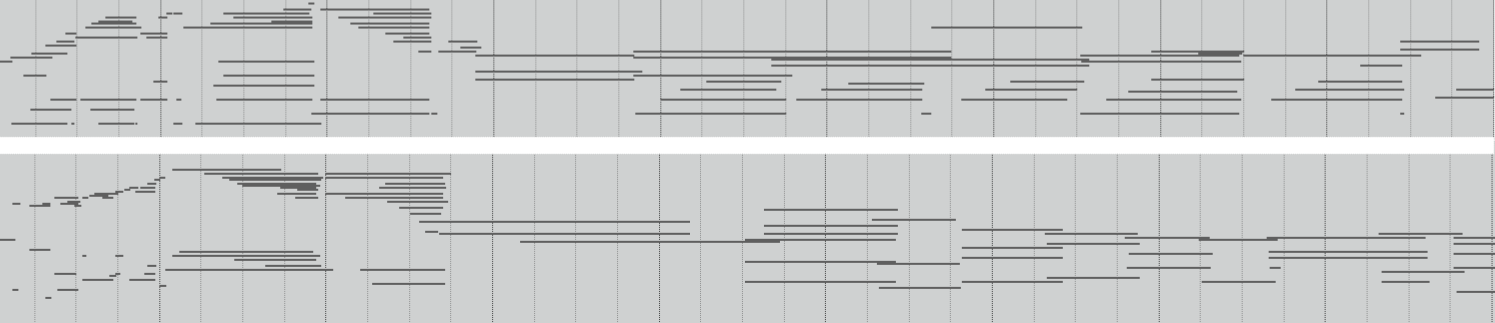}
    \caption{
Two generated piano performances conditioned on the same constraint sequence
using the method described in Sect.~\ref{sec:an-unexp-appl}. The
alternation between fast and slow movements is preserved.
}
  \label{fig:absurdgen}
\end{figure}

\subsubsection{An unexpected application: variations}
\label{sec:an-unexp-appl}
If we ask PIA to generate a whole piece while being conditioned on a complete piano performance (with a constraint sequence containing no $\mathbf{NC}$ symbol), we surprisingly observe that PIA does not copy the constraint sequence, but rather generates a novel piano performance bearing similarities with the constraint sequence (see Fig.~\ref{fig:absurdgen}).

Such unexpected behavior is related to our choice to predict only non constrained tokens as discussed in Sect.~\ref{sec:impr-line-transf-1} and could be used as an additional way to condition generations.
  
\section{PIA, a Max4Live device}
\label{sec:pia-plugin}
\begin{figure}[h]
\centering
\includegraphics[scale=0.2]{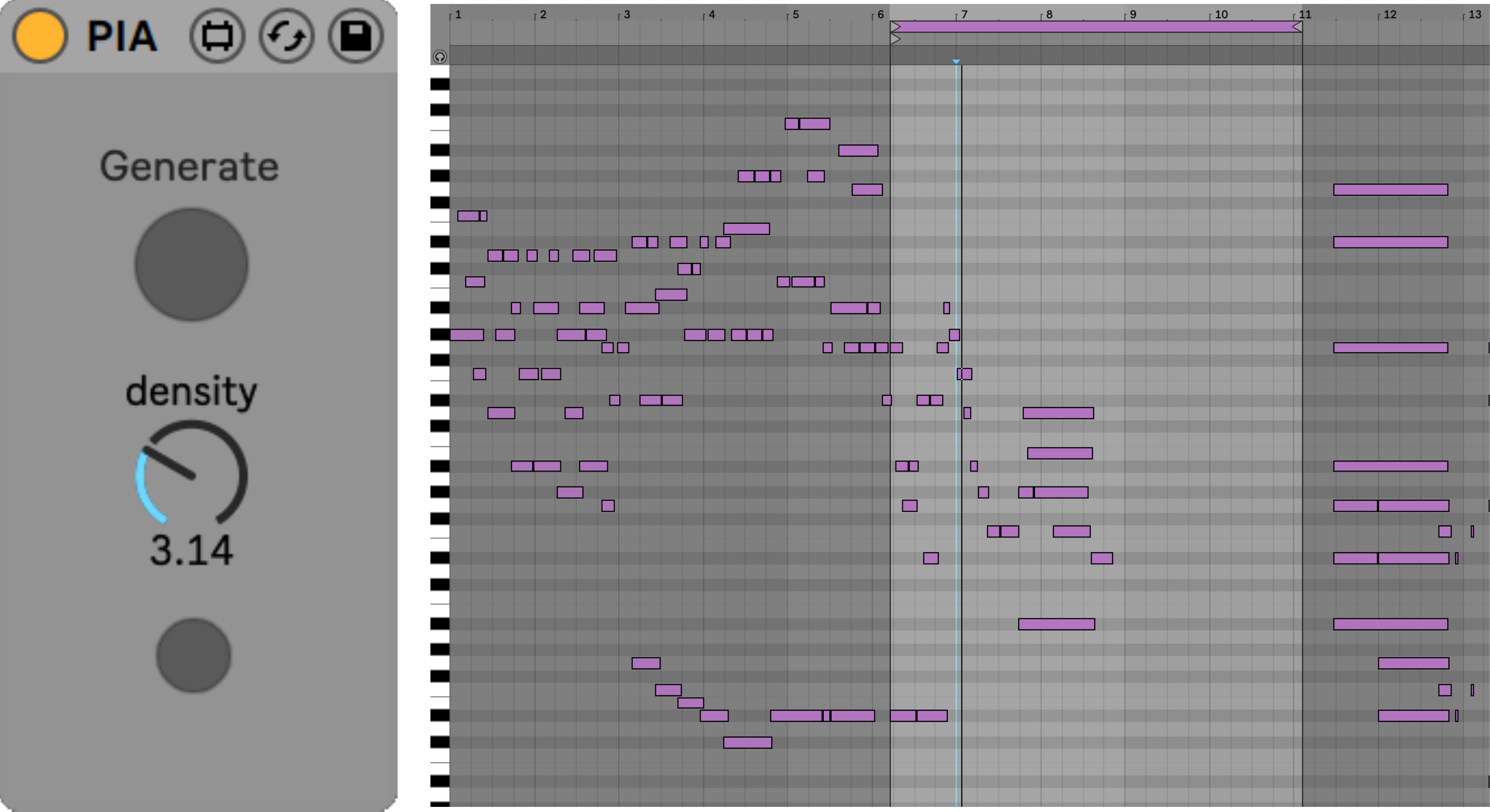} 
    \caption{
The PIA plugin for \emph{Ableton Live} is composed of one button that
triggers the regeneration of a selected region and a slider to choose
the note density.
}
\label{fig:pia}
\end{figure}

We introduce the \textit{PIA} plugin (see Fig.~\ref{fig:pia}) for \emph{Ableton Live}, a
\textit{Max4Live} device able to replace or fill in any contiguous
region of a MIDI pianoroll in \emph{Ableton Live} by relying on the method  presented in
Sect.~\ref{sec:inpa-cont-regi}.
% Flexible
% The architecture we proposed in Sect.~\ref{sec:model} does not impose
% any restriction on the shape or size of the inpainted region. This flexibility allows us to develop an intuitive interaction in \textit{PIA}, where a user simply selects with its mouse the region of a pianoroll to be generated.
% fast

The selected fragment is populated by notes as soon as they are generated.
The first notes start being generated in less than $1$ second, providing a highly-responsive user experience. 
Moreover, users need not wait to listen to the track being generated and generation speed is close to real-time.

  The PIA plugin and a video demonstrating \textit{PIA} in use can be accessed on the
companion website \footnote{\href{https://ghadjeres.github.io/piano-inpainting-application/}{https://ghadjeres.github.io/piano-inpainting-application/}}.

\section{Related Works}
\label{sec:related-works}

\subsection{MIDI encodings}
\label{sec:ss:midi-encoding-rel}
Event-based representations inspired by the \textit{MIDI} format have become the standard to encode MIDI performances \cite{musenet,huang2018music,Oore2020,donahue2019lakhnes}.
In the \textit{MIDI} format, the duration of a note is equal to the cumulative sum of all the time-shift events between the note-on message and its corresponding note-off (see Fig.~\ref{fig:structured-encoding}).
However, as noted in \cite{huang2020pop}, this nested structure embeds an unnecessary complexity in a non-live context, which can hinder the performances of generative algorithms, and carry the risk of generating ill-formed sequences with notes-on event not being properly switched off.

To circumvent that issue, \cite{huang2020pop} proposed a duration-based representation, using the following messages: pitch, velocity, duration and time-shift.
However, they also introduce other types of information such as bar and beat positions and chords labels.
While providing such rich contextual information may improve the generative performances of the model, it also has the disadvantage of ruining the structure of the sequence, as messages corresponding to different musical attributes can occur at any position.
Thus, the predictions have to be made on a single large alphabet enumerating the possible values for all types of information, instead of several smaller alphabets as in our implementation.
For instance, the smallest alphabet found in the literature contains $332$ elements \cite{huang2020pop}, whereas our proposed Structured MIDI encoding typically relies on alphabets of size $88$, $128$ or $106$ (respectively for pitch, velocity or duration and time-shift).

A similar observation was presented in \cite{hsiao2021compound}, but they proposed a solution where the different features of a same note are predicted independently, which we believe weakens the capacity of the model.

\subsection{Models and Interfaces for Interactive Music Generation}
\label{sec:ss:img}
Artificial Intelligence could help improve musicians' workflow by providing an intuitive and convenient way to transform musical ideas into their technical realization.  %which may be otherwise too complex to perform for neophytes.
However, few works jointly design A.I. algorithms for music together with Human-Computer Interfaces to facilitate their usage. Amongst these works, we distinguish two trends: ones that develop standalone interfaces and ones that integrate within existing software.
As an example of the former approach, \cite{Donahue2019} proposed a web-based interface which generates a melody on a 88 keys piano based on a user input limited to a set of 8 keys while \cite{bazin2019nonoto}  developed a model-independent web-based interactive score that can be linked to \emph{Ableton Live} to facilitate musical score inpainting. In \cite{huang2019bach}, the authors proposed a web-based interface for melody reharmonization in the style of Bach. 

Integration within existing software is often approached via the development of plugins: \cite{hadjeres2017deepbach} proposed a plugin for the MuseScore scorewriter editor to perform interactive composition of Bach chorales while
\cite{magentastudio} proposed a series of \textit{Ableton Live} plugins offering multiple ways to generate monophonic melodies. 
 
The aforementioned systems process simpler data compared to piano \textit{MIDI} performances, which limits their expressivity and usages.
%In comparison, PIA can adapt to a relatively wide range of music styles and writing techniques.

%On the other side, as mentioned in the introduction of this paper, powerful autoregressive models able to cope with piano \textit{MIDI} performances offers interactions limited to the continuation of a priming section \cite{musenet,huang2018music,huang2020pop,hsiao2021compound} or harmonization \cite{huang2018music}.
%Our work is an attempt to equip these 

\section{Conclusion}
\label{sec:conclusion}
In this paper, we presented a general architecture for piano inpainting
together with a freely-available \emph{Ableton Live} plugin making it accessible to a
wide audience. We proposed a novel encoding for MIDI piano
performances adapted to this task and an efficient inference scheme, which allowed us to consider moderate-size models and attain almost
real time generation speed while maintaining high generation
capabilities. Through a minimalist user interface, PIA favors rich interactions and suggests
innovative ways to envision music composition; its integration within a
professional DAW advocates for a forthcoming democratization of A.I.-enhanced creation tools.

\section{Acknowledgements}
We thank Adrien Laversanne-Finot for helping us with the implementation of the \textit{Max4Live} device.

\bibliographystyle{abbrv}
\bibliography{pia}
\appendix

\clearpage
\appendixpage

\section{Positional embeddings}
\label{sec:app:embeddings}
Channel embedding is a trainable matrix in $\mathbb{R}^{4 \times 12}$ mapping each of the four channels (pitch, velocity, duration, time-shift) to a 12-dimensional embedding.
Token embedding is a sinusoidal embedding \cite{vaswani2017attention} with dimension $128$ which encodes the position of each token relatively to the beginning of the subsequence being processed by the model.
The elapsed-time embedding is a sinusoidal embedding \cite{vaswani2017attention} of dimension $128$ encoding the absolute time from the beginning of the whole performance. 

The positional embedding is given by the concatenation of the channel, token and elappsed-time embeddings
\begin{equation*}
    p(t) = (p_c(t), p_t(t), p_{e}(t)) \in \mathbb{R}^{268}
\end{equation*}

Both token embedding $p_t$ and elapsed-time embedding $p_e$ are sinusoidal.
The sinusoidal embedding $\mathrm{SE}(t)$ at index $t$ is a $2d$-vector defined by:
\begin{align}
\begin{split}
\mathrm{SE}(t)_{2i} &= \sin (\mathrm{pos}(t) / 10000^{\frac{2i}{d}})\\
\mathrm{SE}(t)_{2i+1} &= \cos (\mathrm{pos}(t) / 10000^{\frac{2i}{d}}),
\end{split}
\end{align}
where $\mathrm{pos}(t) = \lfloor t / 4 \rfloor$ for the token embedding
and $\mathrm{pos}(t) = 100 \sum_{i < \lfloor t / 4 \rfloor} \mathrm{ms}(x^{ts}_i)$ for the elapsed-time embedding, where $\mathrm{ms}(x^{ts}_i)$ maps time shift tokens to their value in milliseconds.

\section{Details of the structured \textit{MIDI} encoding}
\label{sec:app:encoding_alphabebet}
Each attribute is defined on a specific alphabet.
In our implementations, we used the following definitions:
\begin{itemize}
    \item Pitch: $\mathcal{A}_p = \db{21, 108}$ represents the 88 keys  of a piano keyboard.
    \item Velocity: $\mathcal{A}_v = \db{0, 127}$ is a direct legacy from the \textit{MIDI} standard and is sufficient to encodes fine variations in the dynamics.
    \item Duration: $\mathcal{A}_d$ contains $106$ tokens representing an adaptive quantisation of the interval between $0$ and $20$ seconds. See Tab.~\ref{tab:time_spans} for details.
    \item Time-shift: $\mathcal{A}_{ts}$ is defined on the same alphabet as the duration.
\end{itemize}
 
\begin{table}[ht]
\begin{center}
\begin{tabular}{ |c|c|c|c| } 
\hline
& Start (sec.) & End (sec.) & Increment (sec.) \\ 
\hline
Short (50) & $0$ & $0.98$ & $0.02$ \\
\hline
Medium (40) & $1.0$ & $4.9$ & $0.1$ \\
\hline
Long (16) & $5.0$ & $20.0$ & $1.0$ \\
\hline
\end{tabular}
\end{center}
\caption{Time quantisation for duration and time-shift events. 
A time interval ranging from $0$ to $20.0$ seconds is split in $106$ tokens with varying increments over the Short, Medium and Long intervals.
The numbers in parenthesis next to the name of the segments indicate the number of token allocated to each segment.
The values for the start, end and increments values of the different segments are in seconds.
}
\label{tab:time_spans}
\end{table}

\FloatBarrier
\clearpage

\end{document}